\documentclass[proceedings,preprints]{rmaa}

\title{Age and Extinction\\ 
       of the Ultraviolet Emitting Regions in M82}

\author{L.~H. Rodr\'{\i}guez-Merino,\altaffilmark{1}
        D. Rosa-Gonz\'alez,\altaffilmark{1}
        Y.~D. Mayya,\altaffilmark{1}
        L. Carrasco,\altaffilmark{1}
        A. Luna,\altaffilmark{1} \\
        and \\
        R. Romano\altaffilmark{2}}

\altaffiltext{1}{Instituto Nacional de Astrof\'{\i}sica, Optica y Electr\'onica, 
                 Luis Enrique Erro 1, Tonantzintla, Puebla, M\'exico.}
\altaffiltext{2}{Instituto de Astrof\'{\i}sica de Andaluc\'{\i}a, Camino Bajo de 
                 Hu\'etor, 50, Granada, Espa\~na.}

\suppressfulladdresses

\resumen{La galaxia M82 debido a su gran actividad de formaci\'on estelar y a su 
cercan\'{\i}a a la V\'{\i}a Lactea ha sido objeto de m\'ultiples estudios. Empleando 
im\'agenes de M82 que cubren diferentes bandas del espectro electromagn\'etico, 
hemos logrado determinar la edad y extinci\'on de la poblaci\'on estelar localizada 
en regiones con fuerte emisi\'on ultravioleta (UV), estas regiones se encuentran en 
el n\'ucleo y en el disco de M82. Adem\'as, usando las im\'agenes de la emisi\'on UV de 
M82 y las propiedades de sus c\'umulos estelares hemos cuantificado la contribuci\'on 
de \'estos al flujo detectado en esta banda. Encontramos que en regiones nucleares 
la emisi\'on UV es debida a los c\'umulos estelares, mientras que en el disco su 
contribuci\'on disminuye a valores cercanos al 10\%. Los resultados obtenidos nos 
han conducido a inferir que las estrellas de campo del disco de M82 pudieron haber 
sido parte de alg\'un c\'umulo estelar cuando nacieron.}

\abstract{The M82 galaxy has been the subject of several studies basically because 
it is relatively close to to the Milky Way and it displays a strong star formation 
activity. Using multi-band images of M82 we have determined the age and extinction 
of the stellar population located in regions with strong UV emission, these region 
are in the nucleus and the disk of M82. We also have employed the UV images of M82 
and the physical properties of its stellar clusters to measure the contribution of 
the clusters to the detected UV flux. We found that clusters located in the nuclear 
regions are emitting all the observed UV flux, whereas clusters of the disk emit less 
than $\sim$10\%. Based on the results obtained from this work we can infer that the 
field stars located in the disk of M82 could have been part of a stellar cluster when 
they were born.}


\shortauthor{Rodriguez-Merino}
\shorttitle{Star Formation History in M82}

\begin{document}

\maketitle

\section{Introduction.}
The ultraviolet (UV) range of the spectral energy distribution is a well 
known window to perform studies of several kinds of astrophysical phenomena, 
e.g. classifications of celestial sources \citep[see][]{Bianchi07}, star 
formation \citep[see][]{Rosa-Gonzalez07}, galaxy evolution \citep[see][]
{Kaviraj07}, etc.

During the last five years, the {\it Galaxy Evolution Explorer} (GALEX) has been 
retrieving a wealth of information covering the UV interval. This observational 
data can be used to map recent bursts of star formation. It is commonly accepted 
that star clusters born embedded within giant molecular clouds \citep[see][for a 
complete review]{Lada03}, emit strongly in the UV, and form large HII regions. 
The star clusters experience several disrupting processes along their evolution 
which hinder our understanding of the role played by these objects in the galactic 
evolution \citep[see][]{deGrijs07}. The UV images of galaxies that contain a wide 
set of star clusters can be of great help in this direction.

M82 (NGC 3034) is an edge-on spiral galaxy classified recently with a morphological 
SBc type \citep{Mayya05}. It is part of an interacting group of galaxies 
\citep[see][]{Chynoweth08} at a distance of 3.63 Mpc \citep[][]{Freedman94}. The 
interaction between M82 with the biggest member of the group (M81) about one gigayear 
ago brought an intense burst of star formation in the entire disk of M82 \citep{Mayya06}. 
Images of M82 obtained by {\it GALEX} display regions with strong UV emission, which 
are distributed along the disk of the galaxy as well as in the perpendicular direction. 
Recently \citet{Hoopes05} showed that the UV emission aligned with the minor axis is 
the stellar light scattered by dust, and gas which has been shock-heated. However, 
the UV flux detected along the disk is basically light emitted by field stars and star 
clusters. \citet{Mayya08} detected 653 star clusters distributed along the disk of M82, 
some of which are massive star clusters ($\ge10^5 M_{\odot}$). It is possible that at 
least some of the observed disk UV emission is produced by these star clusters.

\begin{figure*}
\centering
\includegraphics[scale=.70]{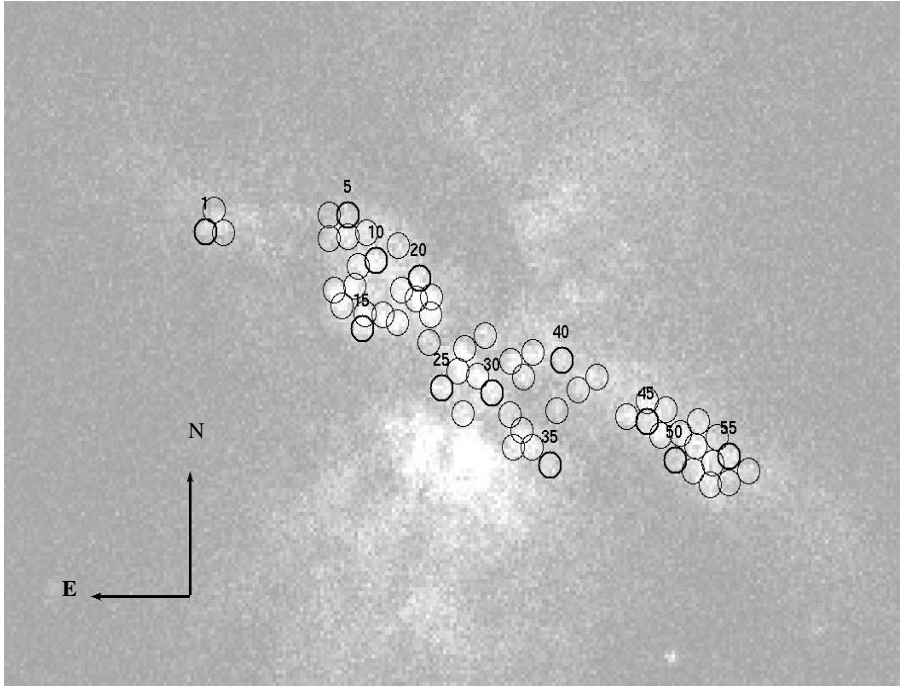}
\caption{This figure displays the position of the 59 apertures defined to study the 
         flux emitted by the nucleus and the disk of M82. Labels correspond to remarked 
         circles.} 
\label{fig:apertures}
\end{figure*}

We use the multi-wavelength data available to investigate the nature of the UV emitting 
regions located in M82. We also estimate the contribution of the star clusters to the UV 
emission of M82. In \S 2 we describe the observational data used. In \S 3 we show the 
results of the multi-wavelength study to trace the age and extinction of the most bright 
regions detected in the far ultraviolet (far-UV) image , and in section \S 4 we discuss 
the fraction contribution  of the star clusters to the observed UV emission. Finally, 
we give our conclusions in \S 5.

\section{Construction of Spectral Energy Distributions}
The data used cover a wide wavelength interval, from the far-UV to the near infrared 
(near-IR). The UV images used in this work are from {\it GALEX} archive and cover 
the far-UV (1350--1750 {\AA}) and near-UV (1750--2750 {\AA}) bands. The optical 
images were obtained from different projects. The image in the U band (3000--4100 
{\AA}) was retrieved from the database of the {\it Sloan Digital Sky Survey} (SDSS), 
whereas the HST/ACS data in the F435, F555 and F814 bands were used as B, V and I 
images. For the near infrared data, we used the J, H and K images obtained with 
the {\it Cananea Near-Infrared Camera} (CANICA) attached to the 2.1 m telescope at 
{\it Observatorio Astron\'omico Guillermo Haro} (OAGH).

The disk of M82 presents large areas with strong UV emission whose nature is yet 
to be understood. We used the far-UV image to define 59 circular apertures distributed 
along the disk. Each aperture is of 5$^{\prime\prime}$ diameter, matching the full 
width half maximum (FWHM) of the point spread function (PSF) of the {\it GALEX}'s 
detector. The selected regions enclose all the areas with strong far-UV emission. 
We point out that some of the bright complexes have several apertures associated 
with it. Figure \ref{fig:apertures} displays the positions of the apertures chosen. 
Aperture numbers between 24--32 and  36--39 correspond to the nuclear starburst.

The zero-points associated with each image were used to convert counts to fluxes 
in order to construct the spectral energy distribution (SED) of the 59 apertures.

\section{Determination of Age and Extinction}
The method used to estimate the age and extinction of the stellar population inside 
each aperture is described below:
\begin{itemize}
\item Theoretical SEDs of simple stellar population (SSP) computed by the Padova 
      Group \footnote{ Kindly provided by A. Bressan.} were employed in this work. 
      The set of theoretical SEDs covers an age interval from 1 Myr to 1 Gyr with 
      variable age step. Models with solar metallicity were used. These models 
      use Kroupa's IMF \citep{Kroupa98}. The flux inside each photometric band was 
      synthesized by integrating the SED over the response curve of the corresponding 
      bands.
\item Each SED was reddened using \citet{Calzetti00} extinction curves with a color 
      excess, E(B-V), running from 0.0 to 2.0 mag, with steps of 0.1 mag.
\item For each aperture a $\chi^2$-type method was used to find the best fit between 
      observed and reddened theoretical SED. The function used to find the best 
      fit was,

      \begin{equation}
         \chi^2 = \sum_{i} (\frac{flux_{i}^{obs}-flux_{i}^{mdl}}{\sigma_{i}^{obs}})^2,
      \end{equation}

      where i= far-UV,..., H, K. $\sigma_{i}^{obs}$ is the uncertainty of observed flux in 
      the i band. The function $\chi^2$ will have the lowest value ($\chi^2_{min}$) for the 
      best match. Currently, we are exploring another method to infer the stellar population 
      properties and its errors.
\end{itemize}

A typical problem related with the determination of stellar parameters using SEDs 
of stellar population is the age-extinction-metallicity degeneracy \citep[see][]
{Worthey94}. However there are several works pointing out that the use of data 
covering a wide wavelength interval is useful to overcome these problems \citep[see]
[]{Bridzius09}.

\begin{figure}[ht]
\centering
\includegraphics[scale=0.40]{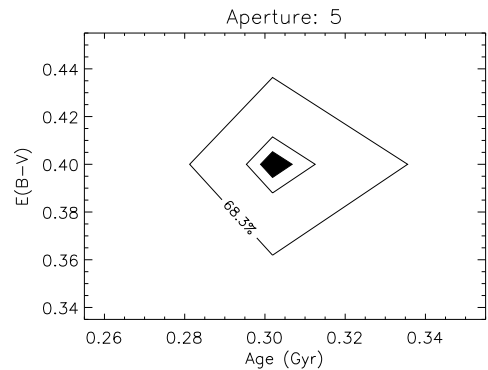}
\caption{The figure displays the results of the $\chi^2$ process used to determine 
         the age and extinction of the stellar population inside the aperture 5. 
         The contour lines enclose age and extinction values of fits with 68, 90, 
         and 95\% of confidence level.}
\label{fig:diag_reg5}
\end{figure}

\begin{figure}[ht]
\centering
\includegraphics[scale=0.40]{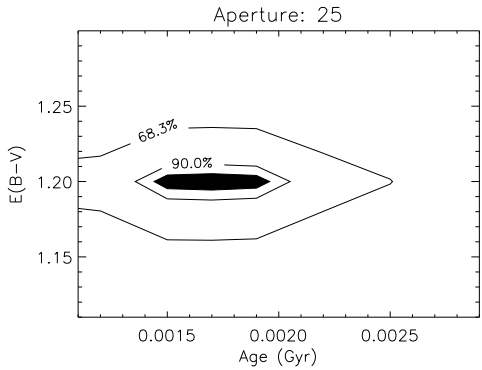}
\caption{Similar to figure \ref{fig:diag_reg5}, but for aperture 25.}
\label{fig:diag_reg25}
\end{figure}

\begin{figure}
\centering
\includegraphics[scale=0.40]{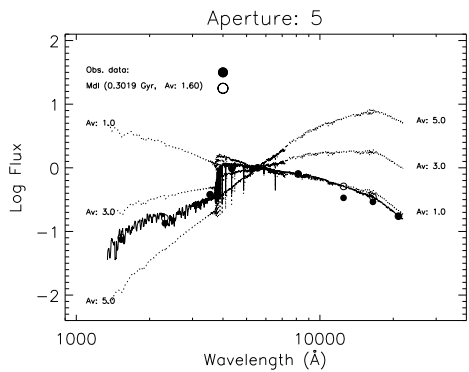}
\caption{The figure displays the fit between the observed spectral energy 
         distribution of aperture 5 and the theoretical SED of an SSP model 
         300 Myr old, reddened with $A_v$= 1.6. A theoretical SED of 10 Myr,  
         reddened with different values of $A_v$ is overplotted (dotted lines) 
         to show that the age-extinction degeneracy is broken with the use of a 
         wide interval in wavelength.}
\label{fig:fit_reg5}
\end{figure}

\begin{figure}
\centering
\includegraphics[scale=0.40]{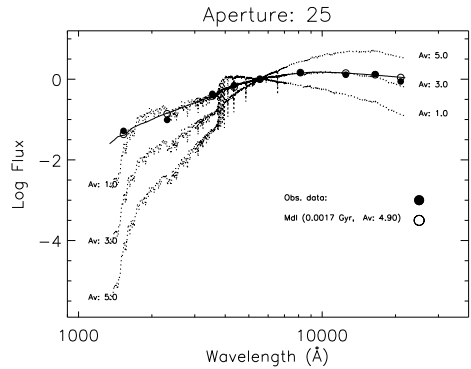}
\caption{The figure displays the fit between the observed spectral energy 
         distribution of aperture 25 and the theoretical SED of an SSP model 1.7 Myr 
         old, reddened with $A_v$= 4.9. A theoretical SED of 500 Myr, 
         reddened with different values of $A_v$ is overplotted (dotted lines).}
\label{fig:fit_reg25}
\end{figure}

We fitted the observed SEDs of all the 59 apertures with SSPs of different age 
and extinction. Because there are more than one SSP that reasonably fits an 
observed SED, we selected not only the best fit model, but also all those SSPs 
with $\chi^2 \le 5\chi^2_{min}$, corresponding to fits at 95\% confidence level.
In figures \ref{fig:diag_reg5} and \ref{fig:diag_reg25}, we display the results 
for apertures 5 and 25 (see Fig. \ref{fig:apertures}). The black area corresponds 
to models at 95\% confidence level. The SED produced by the stellar population 
inside the aperture 5 is best fitted with an SSP model of 300 Myr old, with a 
reddening of E(B-V)=0.4, and the stellar population inside the aperture 25 is best 
fitted with an SSP model of 1.7 Myr old, with a reddening of E(B-V)=1.2. The 
resulting model SEDs are displayed superposed on the observed data in figures 
\ref{fig:fit_reg5} and \ref{fig:fit_reg25}, respectively. Filled circles correspond 
to observed data, circles and continuous line to theoretical model of the best 
fit. Fluxes were normalized to the flux in the V band.

In order to illustrate that the derived age and extinction are not affected 
by degeneracy we have also overplotted the theoretical SED of an SSP for a young 
population (10 Myr) in fig. \ref{fig:fit_reg5}, and relatively older population 
(500 Myr) in fig. \ref{fig:fit_reg25}, in both cases SEDs being reddened with an 
absorption of $A_v$= 1.0, 3.0 and 5.0. It can be seen that in the case of a young 
population (e.g. region 25), the observed SED is not well fitted over the entire 
range of wavelengths with any less reddened model of an old population and the SED 
of an old population (e.g. region 5) is not fitted by a heavily reddened model of 
a young population.

\begin{figure}[t]
\centering
\includegraphics[scale=0.40]{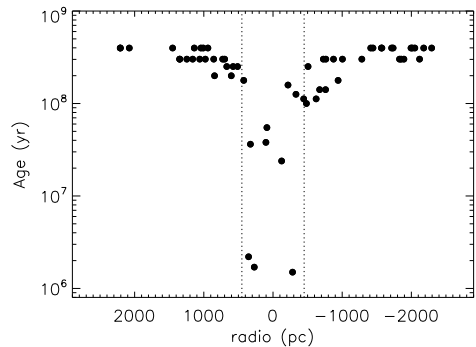}
\caption{The figure shows the age of the stellar population in function of the 
         distance to the galactic center. Positive distances correspond to the 
         north east (NE) and negative distances correspond to the south west (SW). 
         Vertical dotted lines indicates the nuclear area.}
\label{fig:age_radio}
\end{figure}

\begin{figure}[t]
\centering
\includegraphics[scale=0.40]{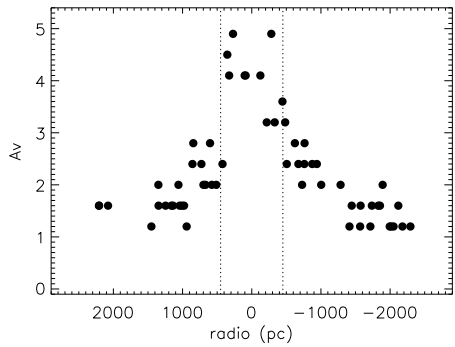}
\caption{The figure shows the extinction in function of the distance to the galactic
         center. The central part of the galaxy is more extincted. Positive distances 
         correspond to NE and negative distances correspond to SW. Vertical dotted
         lines indicates the nuclear area.}
\label{fig:extin_radio}
\end{figure}

Following the method just described, we have determined the age and extinction 
of the stellar population inside each aperture. Figure \ref{fig:age_radio} shows 
the trend followed by the age as a function of the distance to the galactic center. 
It is easily noticeable that all the populations younger than 60 Myr are located 
within a distance of 450 pc from the galactic center. Outside this zone stellar 
populations are older than 100 Myrs, with the oldest UV emitting populations having 
ages of around 400 Myrs. In the zone between 0.5--1.5 Kpc there is an evidence of 
an age gradient. Figure \ref{fig:extin_radio} displays the trend of the 
extinction with the galactocentric radius. In the nuclear area the A$_v$ reaches 
values of $\sim$5.0 mag, decreasing smoothly at larger galactocentric distances.

It is important to stress that we are obtaining age and extinction for the stellar 
population within 59 apertures located in the disk and nuclear parts of M82 using only 
the  photometric data. We were able of constrain the values of both the parameters 
within a small range. Our results are in agreement with the results obtained in 
previous works based on age and extinction sensitive spectroscopic features for 
the disk \citep[see][]{Mayya06} and nucleus \citep[see][]{Forster01}.

\section{UV Emission by Star Clusters}
From the analysis of the previous section, it is clear that the UV flux observed by 
{\it GALEX} in the disk of M82 is contributed entirely from stars. In particular, 
in this galaxy it has been  demonstrated that the disk contains rich population of
compact star clusters \citep[see][]{Mayya08}. So, one of the interesting questions 
is to investigate whether the observed UV emission comes from star clusters or field 
stars. In order to investigate this we have estimated the expected UV fluxes from 
all the clusters within apertures defined in this work. Typically, the disk apertures 
contain less than 10 clusters, whereas nuclear apertures contain more than 10, with 
the aperture 28 containing as much as 80 clusters. We calculated the SED of each 
cluster based on the mass, age, and extinction estimated for each stellar cluster. 
The UV flux in the two {\it GALEX} bands of these SEDs along with the positions of 
the stellar clusters were used to construct two images at the pixel scale of the 
{\it ACS} camera. The high resolution images were convolved with a Gaussian kernel 
of $\sim$5$^{\prime\prime}$ of FWHM and rebined to match the image scale of the 
{\it GALEX} images.

\begin{figure}[t]
\centering
\includegraphics[scale=.50,angle=0]{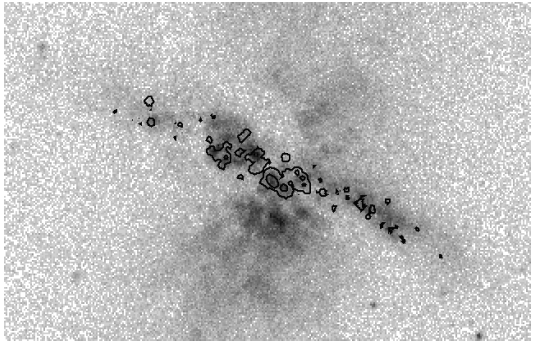}
\caption{This figure displays the far-UV image of M82 obtained by {\it GALEX}, 
         contour lines overplotted correspond to the far-UV synthetic image. 
         The theoretical image was produced with the physical properties of 
         the stellar clusters and the simple stellar populations models.} 
\label{fig:galex_mdl_fuv}
\end{figure}

Figure \ref{fig:galex_mdl_fuv} displays the far-UV image obtained by {\it GALEX} in 
which we have overplotted in contours the synthesized image representing the UV 
contribution of the star clusters only. The simulated image in general reproduces 
the observed large-scale structure in the image. However, on the scale of a few 
parsecs especially in the disk, there are large differences between the two images, 
i.e. there are regions with bright observed emission with negligible contribution 
from the clusters. Similar results were obtained for the image in the near-UV band. 
This could mean that in some regions of the galaxy the field stars are the most 
important sources of ultraviolet flux.

\begin{figure}
\centering
\includegraphics[scale=0.40]{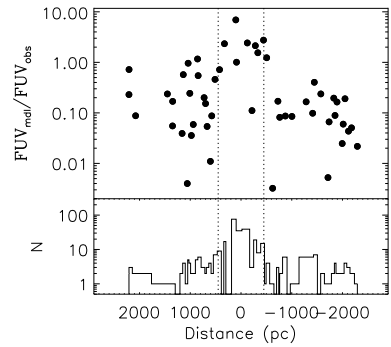}
\caption{The figure displays the comparison between modeled and observed far-UV fluxes as  
         a function to the galactocentric distance. The upper plot shows the ratio between 
         modeled to observed flux, each point corresponds to an aperture, and the lower plot 
         shows the number of star cluster inside the aperture. Positive distances correspond 
         to NE and negative distances correspond to SW. The vertical dotted lines mark the 
         nuclear region.}
\label{fig:fuv_obs_ssp}
\end{figure}

\begin{figure}
\centering
\includegraphics[scale=0.40]{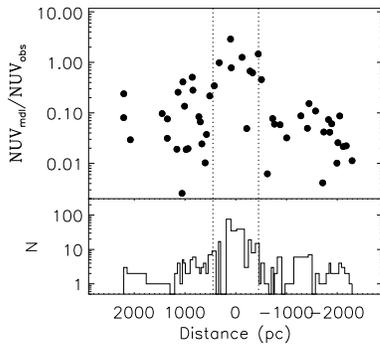}
\caption{Similar to figure \ref{fig:fuv_obs_ssp}, but for the near-UV band.}
\label{fig:nuv_obs_ssp}
\end{figure}

Since the modeled UV flux using only the star cluster properties, does not reproduce 
all the UV flux of the disk detected by {\it GALEX}, it is interesting to estimate 
the percentage of ultraviolet flux emitted by the clusters in all the 59 apertures. 
Figures \ref{fig:fuv_obs_ssp} and \ref{fig:nuv_obs_ssp} display in the upper plots 
the comparison between modeled and observed UV fluxes in the selected apertures. The 
lower plots show the number of star clusters inside each aperture. The two dotted-vertical 
lines enclose all the nuclear regions.

It can be seen that the observed UV flux of majority of the nuclear apertures can be 
completely reproduced by the star clusters, whereas typically only around 10\% of 
the observed emission for the disk apertures can be attributed to the clusters. These 
results show that field stars, not clusters, are the principal contributers to the 
observed UV flux outside the nuclear area.

\citet{Meurer95}, working with the {\it HST} UV images of nine starburst galaxies 
with several bright star clusters, found that 20\% of the total UV luminosity is emitted 
by the clusters. It may be noted that in our work we separate the UV contribution of 
stellar clusters based on their ages. This is uniquely possible in M82 because of the 
spacial segregation of extinguished  young clusters from the not so reddened old 
clusters. Our result that the contribution to the observed UV emission from super star 
clusters (SSCs) is 100\% when young and decreases to $\sim$10\% when they are around 
$10^{8}$ yrs may imply that the clusters in the sample of Meurer are on average of 
intermediate ages. This illustrates the danger of obtaining physical parameters for 
a sample of clusters with a wide distribution of ages.

\section{Conclusions}
We have carried out an analysis of the ultraviolet emission of the nucleus and the disk 
of M82. For this, we used the SEDs covering the far ultraviolet to the near infrared bands 
in 59 apertures of 5$^{\prime\prime}$ diameter.

Using the physical parameters of the star clusters located in M82 and theoretical models 
of simple stellar population, we have estimated the contribution of the star clusters to 
the UV flux detected along the disk of the galaxy. We demonstrate that the star clusters 
located in the nuclear area produce almost all the UV flux, while in the disk the contribution 
of the star clusters is less than 10\%. Given the fact that star clusters in the nuclear 
regions are young ($\sim$10 Myrs) and in the disk are relatively older, the observed 
differences in the relative contribution (100\% in nucleus vs 1--10\% in the disk) imply 
that the cluster contribution to the observed UV flux decreases with age. Most of the UV 
emission from the disk comes from intermediate age field stars and not clusters. The 
synthesized UV image obtained from the parameters of the disk clusters resembles the observed 
image on kiloparsec scales, but not on scales of less than 100 parsecs. This, along with 
the fact that the observed UV emission comes from the disk stars suggests that these stars 
were once part of the clusters, and are now dissolved.


\begin{thebibliography}
\bibitem[Bianchi et al.(2007)]{Bianchi07} Bianchi, L., Rodriguez-Merino, L., et al.\ 
2007, \apjs, 173, 659 

\bibitem[Bridzius et al.(2009)]{Bridzius09} Bridzius, A., Narbutis, D., Stonkute, 
R., Deveikis, V., \& Vansevicius, V.\ 2009, arXiv:0902.3167

\bibitem[Calzetti et al.(2000)]{Calzetti00} Calzetti, D., 
Armus, L., Bohlin, R.~C., Kinney, A.~L., Koornneef, J., \& Storchi-Bergmann, 
T.\ 2000, \apj, 533, 682 

\bibitem[Chynoweth et al.(2008)]{Chynoweth08} Chynoweth, K.~M., 
Langston, G.~I., Yun, M.~S., Lockman, F.~J., Rubin, K.~H.~R., 
\& Scoles, S.~A.\ 2008, \aj, 135, 1983 

\bibitem[de Grijs \& Parmentier(2007)]{deGrijs07} de Grijs, R., \& 
Parmentier, G.\ 2007, Chinese Journal of Astronomy and Astrophysics, 7, 155

\bibitem[F{\"o}rster Schreiber et al.(2001)]{Forster01} 
F{\"o}rster Schreiber, N.~M., Genzel, R., Lutz, D., Kunze, D.,
\& Sternberg, A.\ 2001, \apj, 552, 544

 \bibitem[Freedman et al.(1994)]{Freedman94} Freedman, W.~L., et 
al.\ 1994, \apj, 427, 628

\bibitem[Hoopes et al.(2005)]{Hoopes05} Hoopes, C.~G., et al.\ 
2005, \apj, 619, 99

\bibitem[Kaviraj et al.(2007)]{Kaviraj07} Kaviraj, S., et al.\ 
2007, \apjs, 173, 619

\bibitem[Kroupa(1998)]{Kroupa98} Kroupa, P.\ 1998, Brown Dwarfs 
and Extrasolar Planets, 134, 483 

\bibitem[Lada \& Lada(2003)]{Lada03} Lada, C.~J., \& Lada, 
E.~A.\ 2003, \araa, 41, 57  

\bibitem[Mayya et al.(2008)]{Mayya08} Mayya, Y.~D., 
Romano, R., Rodr{\'{\i}}guez-Merino, L.~H., Luna, A., Carrasco, L., 
\& Rosa-Gonz{\'a}lez, D.\ 2008, \apj, 679, 404

\bibitem[Mayya et al.(2006)]{Mayya06} Mayya, Y.~D., Bressan, 
A., Carrasco, L., \& Hernandez-Martinez, L.\ 2006, \apj, 649, 172 

\bibitem[Mayya et al.(2005)]{Mayya05} Mayya, Y.~D., Carrasco, 
L., \& Luna, A.\ 2005, \apjl, 628, L33 

\bibitem[Meurer et al.(1995)]{Meurer95} Meurer, G.~R., Heckman, 
T.~M., Leitherer, C., Kinney, A., Robert, C., 
\& Garnett, D.~R.\ 1995, \aj, 110, 2665

\bibitem[Rosa-Gonz{\'a}lez et al.(2007)]{Rosa-Gonzalez07} 
Rosa-Gonz{\'a}lez, D., Burgarella, D., Nandra, K., Kunth, D., 
Terlevich, E., \& Terlevich, R.\ 2007, \mnras, 379, 357 

\bibitem[Worthey(1994)]{Worthey94} Worthey, G.\ 1994, \apjs, 
95, 107 
\end{thebibliography}
\end{document}